# SOUND EVENT DETECTION BASED ON AUXILIARY DECODER AND MAXIMUM PROBABILITY AGGREGATION FOR DCASE CHALLENGE 2024 TASK 4

## Technical Report

*Sang Won Son[1], Jongyeon Park[1], Hong Kook Kim[1,2]*   *Sulaiman Vesal[3]*   *Jeong Eun Lim[4]*

[1] AI Graduate School,
[2] School of EECS
Gwangju Institute of Science and Technology
Gwangju 61005, Korea
{{ssw970519, jypark3737}@gm., hongkook@}gist.ac.kr

[3] AI Lab., Innovation Center
Hanwha Vision
Teaneck, New Jersey
07666, USA
s.vesal@hanwha.com

[4] AI Lab., R&D Center
Hanwha Vision
Seongnam-si, Gyeonggi-do
13488, Korea
je04.lim@hanwha.com

**ABSTRACT**

In this report, we propose three novel methods for developing a sound event detection (SED) model for the DCASE 2024 Challenge Task 4. First, we propose an auxiliary decoder attached to the final convolutional block to improve feature extraction capabilities while reducing dependency on embeddings from pre-trained large models. The proposed auxiliary decoder operates independently from the main decoder, enhancing performance of the convolutional block during the initial training stages by assigning a different weight strategy between main and auxiliary decoder losses. Next, to address the time interval issue between the DESED and MAESTRO datasets, we propose maximum probability aggregation (MPA) during the training step. The proposed MPA method enables the model's output to be aligned with soft labels of 1 s in the MAESTRO dataset. Finally, we propose a multi-channel input feature that employs various versions of logmel and MFCC features to generate time-frequency pattern. The experimental results demonstrate the efficacy of these proposed methods in a view of improving SED performance by achieving a balanced enhancement across different datasets and label types. Ultimately, this approach presents a significant step forward in developing more robust and flexible SED models.

*Index Terms*—Sound event detection (SED), semi-supervised learning, auxiliary decoder, multi-channel input feature, maximum probability aggregation

## 1. INTRODUCTION

The objective of sound event detection (SED) is to recognize and classify individual sound events originating from acoustic signals, along with their corresponding time stamps. In recent years, SED has been extensively researched using deep learning models [1]. However, a significant challenge when using deep learning for SED is the requirement of strong labels, which are both expensive and time-consuming. This problem has resulted in research on developing weakly supervised and semi-supervised learning techniques, as well as the use of soft labels with a large time interval instead of strong labels with a short time interval.

During DCASE 2023 Challenge Task 4A, we proposed a convolutional recurrent neural network (CRNN) [2] that had convolution blocks composed of frequency dynamic convolution [3] and large kernel attention [4] blocks. Thus, we achieved first place in terms of PSDS Scenarios 1 (PSDS1) and 2 (PSDS2) [5] with scores of 0.591 and 0.835, respectively, for the evaluation dataset. Nevertheless, there were two problems when implementing the system in this year challenge. Due to the embeddings extracted from pretrained large models, such as bidirectional encoder representations from audio transformers (BEATs) [6], the performance of convolutional blocks with learnable parameters during training can be limited. Furthermore, the soft labels in the MAESTRO dataset [7] are recorded in segments of 1 s, unlike the strong labels in the DESED dataset [8].

In this submission, we aim to improve the SED model [2] by proposing three novel methods. First, to enhance the ability of the convolutional block to extract features, we apply an auxiliary decoder attached to the last convolutional block. The proposed auxiliary decoder enhances the feature extraction capabilities of the convolutional block while reducing the dependency on embeddings extracted by the BEATs encoder. Second, with regard to different time intervals, maximum probability aggregation (MPA) is proposed to ensure that the output of the MAESTRO dataset matches the time interval between the soft label and the model's output. Finally, to extract useful features for the DESED and MAESTRO datasets, we propose a multi-channel input feature that employs various versions of logmel and MFCC features to generate time-frequency pattern.

Following this introduction, Section 2 presents a description of the dataset and input features of the SED model developed in this study. Section 3 proposes the auxiliary decoder, MPA, and multi-channel input feature. Section 4 evaluates the performance of the submitted SED model on the DCASE 2024 Task 4 validation dataset. Finally, Section 5 concludes this report.

---
* This work was supported in part by Hanhwa Vision Co. Ltd., and by the Institute of Information & communications Technology Planning & Evaluation(IITP) grant funded by the Korean government (MSIT) (No.2022-0-00963, Localization Technology Development on Spoken Language Synthesis and Translation of OTT Media Contents).

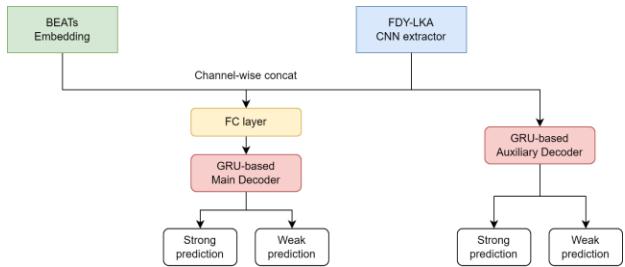

Figure 1. Illustration of the proposed auxiliary decoder to enhance performance of convolution layers.

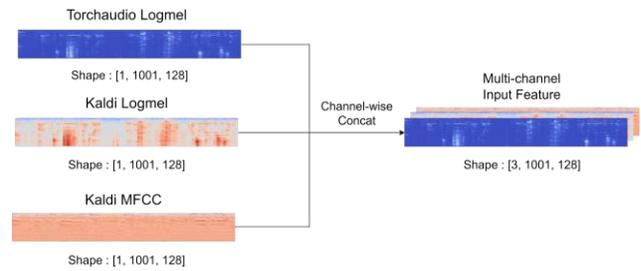

Figure 2. Diagram of the proposed multi-channel input feature for SED system, where the Torchaudio framework is used to generate these features.

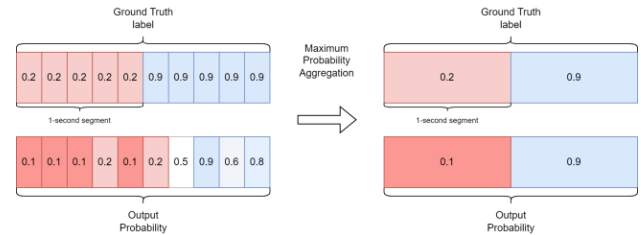

Figure 3. Diagram of the proposed maximum probability aggregation that is applied only for the MAESTRO dataset during training.

## 2. DATASET

Unlike 2023, the datasets for DCASE 2024 Challenge Task 4 can be divided into two categories: the DESED and MAESTRO Real datasets. There are several types of data in the DESED datasets: weakly labeled data, unlabeled in-domain training data, strongly labeled synthetic data, and strongly-labeled real data. The strongly labeled synthetic dataset is distinct from the other datasets by being created by Scraper [9]. The weakly labeled dataset only contains class labels and is annotated for 1,578 clips. The unlabeled in-domain training dataset contains 14,412 audio clips. Finally, the real strongly labeled and synthetic datasets contain 3,470 and 10,000 clips, respectively.

The MAESTRO Real dataset originally contains longer audio clips that lasted more than 180 s. However, to use both datasets in a similar condition, the audio clips are cropped to 10 s, allowing an intersection of 9 s. In the MAESTRO real dataset, there are soft labels recorded in segments of 1 s for 17 classes. In this challenge, we combine similar classes in the DESED and MASETRO datasets as one class, for example, "Speech" in the DESED dataset and "People Talking" in the MAESTRO dataset, and "Dishes" in the DESED dataset and "Cutlery and dishes" in the MAESTRO dataset.

In the following preprocessing procedures, the provided data are used as the input to an SED model, starting by resampling the mono-channel signals from 44.1 to 16 kHz. Then, the audio signals are split into frames of 2,048 samples, each with a hop length of 160 samples. Each frame first performs a 2,048-point fast Fourier transform (FFT) employing 2,048 points, followed by a 128-dimensional mel-filterbank analysis. There are 1,001 frames for each 10-s audio clip. As a result, the input feature dimensions are $1001 \times 128$. The retrieved mel-spectrogram features are then normalized using the mean and standard deviation for all training audio samples. When extracting multi-channel input feature, we use identical parameters for preprocessing.

## 3. PROPOSED METHOD

In this challenge, our network architecture is identical to that in [2], which was proposed in the previous challenge. Rather than improving the architecture, we focus on a learning strategy to make better performance on the MAESTRO dataset, while mitigating performance degradation on the DESED dataset.

### 3.1. Auxiliary decoder

To improve the performance of convolutional blocks, we use an auxiliary decoder attached to the last convolutional block, as shown in Fig. 1. This auxiliary decoder has an identical structure to the main decoder: two bidirectional gated recurrent units (Bi-GRUs) designed to learn temporal context information and a fully connected classifier with a sigmoid function to generate the probability of each class. It should be noted that the auxiliary decoder does not share weights with the GRU-based main decoder. The auxiliary decoder is only used for training step. In other words, only the main decoder is activated to make output of the SED model during test step. In particular, we assign a higher weight to the loss from the auxiliary decoder's output during the early stages of training to further enhance the performance of the convolutional block.

### 3.2. Multi-channel logmel feature

This year challenge has two different datasets such as DESED and MAESTRO Real dataset and their target classes are different. Thus, the logmel features in the baseline of the last year are also different depending on the datasets. To accomodate the gap between the two different datasets, we propose to use multi-channel logmel feature that is obtained by several settings of logmel extraction. Specifically, we use three different logmel-based features: a logmel spectrogram extracted by the Torchaudio framework, one extracted by kaldi in the Torchaudio framework [10], and the MFCC feature by using kaldi in the framework. After generating these input features, they are concatenated channel-wise to create the multi-channel logmel feature. After concatenation, the input feature becomes the input of our SED model for training and testing the model.

### 3.3. Length-adjustable maximum probability aggregation

Table 1: Performance comparison of the baseline and various versions of the proposed SED model on the validation and public evaluation datasets of the DCASE 2024 Challenge Task 4.

| Model | Auxiliary decoder | Maximum probability aggregation | Multi-channel input feature | Ensemble | Validation Dataset | | Public Evaluation Dataset |
|---|---|---|---|---|---|---|---|
| | | | | | Class score-based PSDS | Macro average pAUC-score | Class score-based PSDS |
| Baseline: CRNN-based mean-teacher model [19] | – | – | – | – | 0.49 ± 0.004 | 0.73 ± 0.007 | 0.5502 |
| FDY–LKA-CRNNv0 | – | – | – | – | 0.472 ± 0.003 | 0.685 ± 0.015 | 0.568 ± 0.004 |
| FDY–LKA-CRNNv1 | – | √ | – | – | 0.479 ± 0.003 | 0.690 ± 0.015 | 0.575 ± 0.026 |
| FDY–LKA-CRNNv2 | √ | √ | – | – | 0.488 ± 0.002 | 0.681 ± 0.013 | 0.588 ± 0.053 |
| FDY–LKA-CRNNv3 | – | √ | √ | – | 0.481 ± 0.004 | 0.716 ± 0.024 | 0.547 ± 0.024 |
| FDY–LKA-CRNNv4 | √ | √ | √ | – | 0.486 ± 0.008 | 0.694 ± 0.012 | 0.586 ± 0.012 |
| FDY–LKA-CRNNv5 | √ | √ | √ | √ | 0.485 ± 0.005 | 0.699 ± 0.006 | 0.585 ± 0.006 |

There are different labeling and time intervals between two datasets. That is, strong labels are assigned once every 10-ms segment in the DESED dataset, whereas soft labels of MAESTRO dataset are assigned once every 1-s segment. However, this causes a mismatch between the label information and the output of the model because the model provides the estimated labels once every 40 ms. Moreover, an event in the MAESTRO dataset that is soft-labeled on a 1-s segment does not always exist in a whole second when the event is required to be hard-labeled. To mitigate this problem, we propose MPA based on 1 s only for the MAESTRO dataset to match the time interval between the output of the SED model and the soft labels. Since a segment in the SED model is composed of 25 frames which corresponds to 1 s. Therefore, we transform the highest probability value among 25 frames per class into the class probability for a 1-s segment for the MAESTRO dataset, which results in an identical time interval to the soft labels. Notice that this MPA is only applied during the training step.

## 4. EXPERIMENTAL RESULTS

### 4.1. Model training

The parameters of the FDY–LKA-CRNN-based SED model were initialized in the training step using Xavier initialization [10]. The minibatch-wise adaptive moment estimation optimization technique [11] was employed, which involved decoupling the weight decay from the gradient-based updates. Additionally, a dropout method [12] was applied to the FDY–LKA-CRNN model at a rate of 0.5. The learning rate was set according to the ramp-up strategy [13], with the maximum learning rate reaching 0.001 after 50 epochs. Several augmentation techniques were applied to the train data, including time-frequency shift [14], time mask [15], mix-up [16], and filter augmentation [17].

### 4.2. Discussion

The performance of the proposed SED model was evaluated using the measures defined in the DCASE 2024 Challenge Task 4 [18], such as macro-average pAUC (MPAUC) and PSDS. Table 1 compares the performance between the baseline and various versions of the proposed SED models on the validation and public evaluation datasets of the DCASE 2024 Challenge Task 4. As presented in the table, the proposed FDY–LKA-CRNN-based mean teacher model with MPA (FDY–LKA-CRNNv1) on the validation dataset showed improvement of 0.007 and 0.005 in the PSDS and MPAUC, repectively, than the FDY-LKA-CRNNv0 submitted for DCASE 2023 challenge Task 4A. In addition, applying the auxiliary decoder (FDY–LKA-CRNNv2) demonstrated superior performance, with increases in PSDS of 0.016 and decrease in MPAUC of 0.004, respectively, compared to FDY–LKA-CRNNv0. In comparison, the multi-channel input feature was combined with MPA (FDY–LKA-CRNNv3) achieved higher MPAUC score of 0.031 than FDY–LKA-CRNNv0. Next, combining all the proposed methods yielded higher PSDS and MPAUC of 0.014 and 0.011, respectively, than FDY–LKA-CRNNv0. Finally, ensembling 64 models obtained from different checkpoints during training FDY–LKA-CRNNv4 scored 0.013 higher in PSDS and 0.014 higher in MPAUC, compared to FDY–LKA-CRNNv0.

We repeated performance evaluation on the public evaluation dataset of the DCASE 2024 Challenge Task 4. In the experiment, all the versions of the FDY–LKA-CRNN-based SED model (FDY-LKA-CRNNv0) scored PSDS higher than the Baseline system. Among the single models, FDY–LKA-CRNNv2 showed superior increases in PSDS of 0.020, compared to FDY–LKA-CRNNv0. An ensemble model, FDY–LKA-CRNNv5, showed scored 0.017 higher in PSDS, compared to FDY–LKA-CRNNv0.

## 5. CONCLUSION

In this report, we proposed maximum probability aggregation and an auxiliary decoder to improve SED performace for DCASE 2024 Challenge Task 4. The proposed auxiliary decoder was employed during the training step. Subsequently, the auxiliary decoder enhanced the performance of the convolutional block, enabling it to extract semantics. MPA was applied to MAESTRO dataset to match the time alignment between the output of SED model and soft labels. In addition, we proposed multi-channel input feature to accommodate the various time-frequency patterns from the two different datasets used in this challenge. The experimental results showed that our proposed methods employed to FDY-LKA-CRNN based SED model improved metric on validation dataset of the DCASE 2024 Challenge Task 4. The performance of various ensemble models derived from the model checkpoints was also explored. The results demonstrated that an ensemble model comprising the Top 1–64 models achieved a 0.075 and 0.0248 increases in PSDS and MPAUC about validation datsaset, respectively, compared to FDY–LKA-CRNNv0. Furthermore, it also scored 0.017 higher in PSDS than FDY–LKA-CRNNv0 on the public evaluation dataset.